\documentclass[aps, prl, 10pt, twocolumn, showpacs,
preprintnumbers, superscriptaddress, amsmath, amssymb]{revtex4-1}
\usepackage[pdftex]{graphicx}
\usepackage[ngerman,english]{babel}
\usepackage[ansinew]{inputenc}
\usepackage{natbib}
\usepackage{color}
\usepackage{siunitx}
\usepackage{calc}

\definecolor{green2}{rgb}{.0, .58, 0}

\begin{document}

\title{Vectorial scanning force microscopy using a nanowire sensor}

\author{N. Rossi}
\affiliation{Department of Physics, University of Basel, Klingelbergstrasse 82, 4056 Basel, Switzerland}

\author{F. R. Braakman}
\affiliation{Department of Physics, University of Basel, Klingelbergstrasse 82, 4056 Basel, Switzerland}

\author{D. Cadeddu}
\affiliation{Department of Physics, University of Basel, Klingelbergstrasse 82, 4056 Basel, Switzerland}

\author{D. Vasyukov}
\affiliation{Department of Physics, University of Basel, Klingelbergstrasse 82, 4056 Basel, Switzerland}

\author{G. T\"ut\"unc\"uoglu}
\affiliation{Laboratiore des Mat\'{e}riaux Semiconducteurs, Institut des Mat\'{e}riaux, \'{E}cole Polytechnique F\'{e}d\'{e}rale de Lausanne, 1015 Lausanne, Switzerland}

\author{A. Fontcuberta i
Morral}
\affiliation{Laboratiore des Mat\'{e}riaux Semiconducteurs, Institut des Mat\'{e}riaux, \'{E}cole Polytechnique F\'{e}d\'{e}rale de Lausanne, 1015 Lausanne, Switzerland}

\author{M. Poggio}
\affiliation{Department of Physics, University of Basel, Klingelbergstrasse 82, 4056 Basel, Switzerland}

\date{\today}

\begin{abstract}Self-assembled nanowire (NW) crystals can be grown
  into nearly defect-free nanomechanical resonators with exceptional
  properties, including small motional mass, high resonant frequency,
  and low dissipation.  Furthermore, by virtue of slight asymmetries
  in geometry, a NW's flexural modes are split into doublets
  oscillating along orthogonal axes.  These characteristics make
  bottom-up grown NWs extremely sensitive vectorial force sensors.
  Here, taking advantage of its adaptability as a scanning probe, we
  use a single NW to image a sample surface.  By monitoring the
  frequency shift and direction of oscillation of both modes as we
  scan above the surface, we construct a map of all spatial tip-sample
  force derivatives in the plane.  Finally, we use the NW to image
  electric force fields distinguishing between forces arising from the
  NW charge and polarizability.  This universally applicable technique
  enables a form of atomic force microscopy particularly suited to
  mapping the size and direction of weak tip-sample forces.
\end{abstract}


\maketitle

\section{Introduction}
Atomic force microscopy
(AFM)~\cite{binnig_atomic_1986,giessibl_advances_2003} exists in
several forms and is now routinely used to image a wide variety of
surfaces, in some cases with atomic~\cite{giessibl_atomic_1995} or
sub-atomic resolution~\cite{giessibl_subatomic_2000}. Due to its
versatility, this technique has found application in fields including
solid-state physics, materials science, biology, and medicine.
Variations on the basic technique, including contact and non-contact
modes, allow its application under diverse conditions and with
enhanced contrast for specific target signals.  The measurement of
multiple mechanical harmonics yields information on the non-linearity
of the tip-sample interaction, while monitoring higher mechanical
modes provides additional types of imaging contrast.  Today, these
various types of AFM are most often carried out using cantilevers
processed by top-down methods from crystalline Si, which are hundreds
of $\mu$m long, tens of $\mu$m wide, and on the order of one $\mu$m
thick.

In recent years, researchers have developed new types of mechanical
transducers, fabricated by bottom-up
processes~\cite{poggio_sensing_2013}.  These resonators are built
molecule-by-molecule in processes that are typically driven by
self-assembly or directed self-assembly. Prominent examples include
doubly-clamped carbon nanotubes (CNTs)~\cite{sazonova_tunable_2004},
suspended graphene sheets~\cite{bunch_electromechanical_2007}, and
nanowire (NW)
cantilevers~\cite{perisanu_high_2007,feng_very_2007,nichol_displacement_2008,li_bottom-up_2008,belov_mechanical_2008,
  gil-santos_nanomechanical_2010}.  Assembly from the bottom up allows
for structures with extremely small masses and low defect densities.
Small motional mass both enables the detection of atomic-scale
adsorbates and results in high mechanical resonance frequencies,
decoupling the resonators from common sources of noise.  Near
structural perfection results in low mechanical dissipation and
therefore high thermally-limited force sensitivity.  These factors
result in extremely sensitive mechanical sensors: e.g.\ CNTs have
demonstrated yg mass resolution~\cite{chaste_nanomechanical_2012} and
a force sensitivity close to 10 zN/$\sqrt{\text{Hz}}$ at cryogenic
temperatures~\cite{moser_ultrasensitive_2013}.

Nevertheless, given their extreme aspect ratios and their ultra-soft
spring constants, both CNTs and graphene resonators are extremely
difficult to apply in scanning probe applications.  NWs on the other
hand, when arranged in the pendulum geometry, i.e.\ with their long
axis perpendicular to the sample surface, are well-suited as scanning
probes, with the pendulum geometry preventing the tip from snapping
into contact~\cite{gysin_low_2011}.  When approached to a surface, NWs
experience extremely low non-contact friction making possible
near-surface ($< 100$ nm) force sensitivities around 1
aN/$\sqrt{\text{Hz}}$~\cite{nichol_nanomechanical_2012}.  As a result,
NWs have been used as force transducers in nuclear magnetic resonance
force microscopy~\cite{nichol_nanoscale_2013} and may be amenable to
other ultra-sensitive microscopies such as Kelvin probe force
microscopy~\cite{nonnenmacher_kelvin_1991} or for the spectroscopy of
small friction forces~\cite{stipe_noncontact_2001}.  Furthermore,
their highly symmetric cross-section results in orthogonal flexural
mode doublets that are nearly
degenerate~\cite{nichol_displacement_2008,gil-santos_nanomechanical_2010}.
In the pendulum geometry, these modes can be used for the simultaneous
detection of in-plane forces and spatial force derivatives along two
orthogonal directions~\cite{gloppe_bidimensional_2014}.  Although
one-dimensional (1D) dynamic lateral force microscopy can be realized
using the torsional mode of conventional AFM
cantilevers~\cite{pfeiffer_lateral-force_2002,giessibl_friction_2002,kawai_dynamic_2005,kawai_direct_2009,kawai_ultrasensitive_2010},
the ability to simultaneously image all vectorial components of
nanoscale force fields is of great interest.  Not only would it
provide more information on tip-sample interactions, but it would also
enable the investigation of inherently 2D effects, such as the
anisotropy or non-conservative character of specific interaction
forces.

Here, we use an individual as-grown NW to realize the vectorial
scanning force microscopy of a patterned surface.  By monitoring the
NW's first-order flexural mode doublet, we fully determine the
magnitude and direction of the static tip-sample force derivatives in
the 2D scanning plane. As a proof-of-principle, we also map the force
field generated by voltages applied to a sample with multi-edged gate
electrodes and identify the contributions of NW charge and
polarizability to sample-tip interactions.

\section{Nanowire force sensors}
The GaAs/AlGaAs NWs studied here are grown by molecular beam epitaxy
(MBE), as described in the Appendix.  They have a predominantly
zinc-blende crystalline structure and display a regular hexagonal
cross-section.  The NWs are grown perpendicular to the Si growth
substrate and remain attached to it during the measurements in order
to maintain good mechanical clamping and to avoid the introduction of
defects through processing.  Both these factors help to minimize
mechanical dissipation.  Measurements are performed with the NWs
enclosed in a UHV chamber at the bottom of a liquid $^{4}\text{He}$
bath cryostat with a base temperature of 4.2 K and pressure of
$10^{-7}$ mbar.  The experimental setup allows us to
interferometrically monitor the displacement of single NWs as we scan
them above a sample surface, as shown in Fig.~1(a) and as described in
the Appendix.  In the presented measurements, we use two individual
NWs, one of which is shown in the scanning electron micrograph (SEM)
in Fig.~1(b), however similar results were obtained from several other
NWs.  We can optionally drive the mechanical motion of the NW using a
piezoelectric transducer fixed to the back of the NW chip holder.

The NWs can be characterized by their displacement noise spectral
density.  Here, the measured mechanical response is driven only by the
Langevin force resulting from the coupling between the NW and the
thermal bath.  As shown in Fig.~1(d), NW1 shows two distinct resonance
peaks at $f_1 = 414$ kHz and $f_2 = 420$ kHz, corresponding to the two
fundamental flexural eigenmodes polarized along two orthogonal
directions (see schematic in Fig.~1(c)).  The modes are split by
$\delta = 6.26$ kHz and have nearly identical quality factors of $Q_i
= 5 \times 10^4$, as determined by both ring-down measurements and by
fitting the thermal noise spectral density with that of a damped
harmonic oscillator.  The splitting between the modes is many times
their linewidths, a property observed in several measured NWs.  Finite
element modeling (FEM) has shown that even small cross-sectional
asymmetries ($<1\%$) or clamping asymmetries can lead to mode
splittings similar to those
observed~\cite{cadeddu_time-resolved_2016}.

\begin{figure}[t]
  \centering
  \includegraphics[width=1\columnwidth]{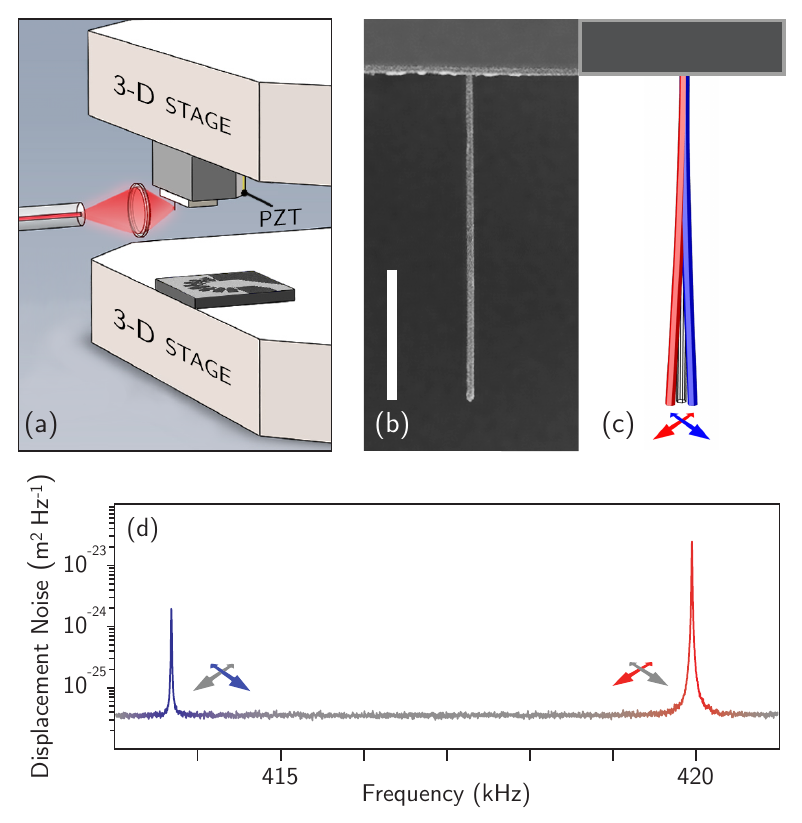}
\caption{
  \label{fig1}
  \textbf{Schematic drawing of the measurement setup}.  \textbf{a} A
  fiber-optic interferometer is aligned with a single NW using a
  piezoelectric positioning stage (top).  A second stage (bottom) is
  used to position and scan the sample surface under the NW.
  \textbf{b}, A scanning electron micrograph (SEM) of NW1, a typical
  GaAs/AlGaAs NW.  The scale bar represents 10 $\mu$m. \textbf{c}, A
  schematic diagram showing the two orthogonal fundamental flexural
  modes of the NW.  \textbf{d} The displacement spectral noise density
  of the fundamental mode doublet measured by fiber-optic
  interferometry for NW1.  We extract $\theta_0 = 15^\circ$. }
\end{figure}

We resolve the two first-order flexural modes with different
signal-to-noise ratio, given that the principal axes
$\mathbf{\hat{r}}_1$ and $\mathbf{\hat{r}}_2$ of the modes are rotated
by some angle $\theta_0$ with respect to the optical detection axis
$\mathbf{\hat{x}}$~\cite{nichol_displacement_2008}.  The total
measured displacement is $x(t) =
r_1(t)\sin\theta_0+r_2(t)\cos\theta_0$, where $r_1$ and $r_2$
represent the displacement of each flexural mode.  The mean square
displacement generated by uncorrelated thermal noise is then $\langle
x^2\rangle = P_1 + P_2$, where $P_1 = \langle r^2_1 \rangle
\sin^2\theta_0$ and $P_2 = \langle r^2_2 \rangle \cos^2\theta_0$
represent the integrated power of each measured resonance in the
spectral density.  Given that the motional mass $m$ of the two
orthogonal flexural modes is the same, using the equipartition
theorem, we find the ratio of their mean-square thermal displacements
$\langle r_1^2 \rangle / \langle r_2^2 \rangle = f_2^2 / f_1^2$.
Therefore, from the measured thermal peaks in the spectral density, we
calculate the angle $\theta_0 = \arctan \left (\frac{f_1}{f_2}
  \sqrt{\frac{P_{1}}{P_{2}}} \right )$ between $\mathbf{\hat{r}}_1$
($\mathbf{\hat{r}}_2$) and $\mathbf{\hat{x}}$ ($\mathbf{\hat{y}}$).
Furthermore, since $k_i = k_B T / \langle r_i^2 \rangle$, where $i =
1,2$, we obtain the spring constants of each flexural mode, which are
typically on the order of 100 mN/m.  These parameters yield mechanical
dissipations $\Gamma_i = k_i / (2 \pi f_i Q_i)$ and thermally limited
force sensitivities $S_{F_i}^{1/2} = \sqrt{4 k_B T \Gamma_i}$ around
10 ng/s and 5 aN/$\sqrt{\text{Hz}}$, respectively.

\begin{figure}[t]
  \centering
  \includegraphics[width=1\columnwidth]{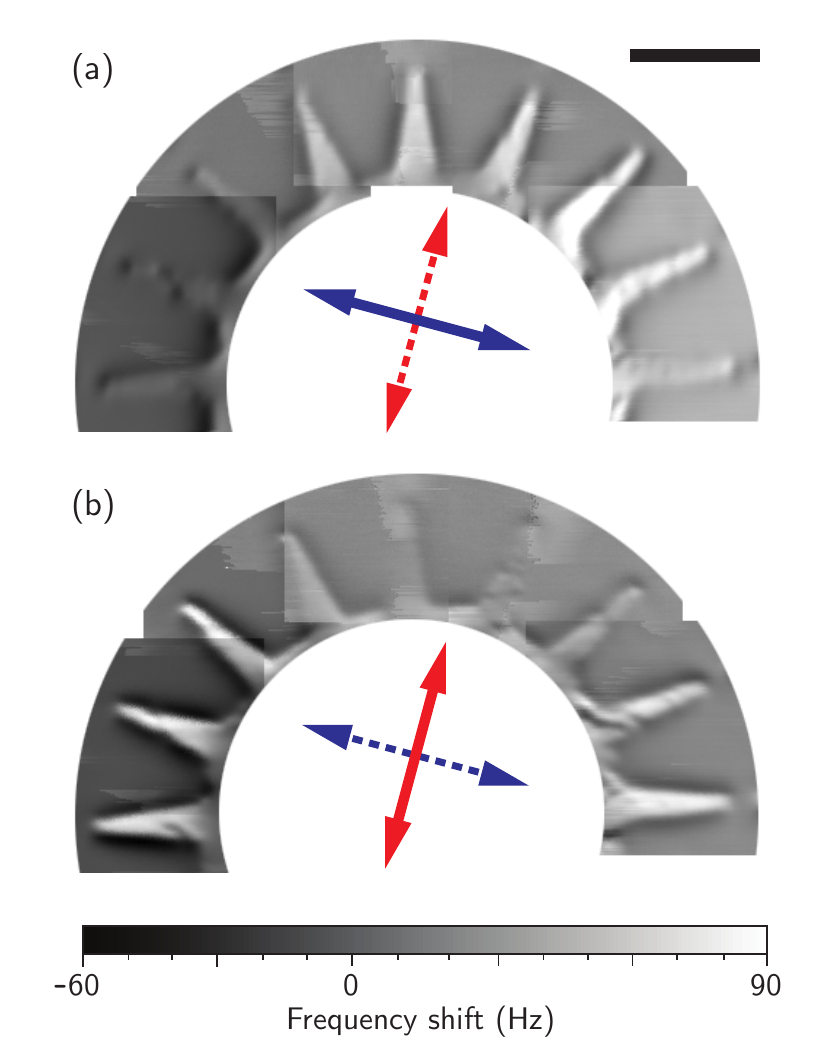}
\caption{
\label{fig2}
\textbf{Mode 1 and mode 2 frequency shift images}. Left (Right) panel
shows $\Delta f_1(x,y)$ ($\Delta f_2(x,y)$) of NW1, , which has
$\theta_0 = 15^\circ$, at a distance of 70 nm from a pattern of Au
finger gates on a Si substrate.  Gray scale is proportional to the
respective frequency shifts and the blue (red) arrows show the
direction $\mathbf{\hat{r}}_1$ ($\mathbf{\hat{r}}_2$) of mode 1 (2) as
determined by thermal noise measurements far from the surface.  The
scale bar represents 2 $\mu$m.}
\end{figure}

\section{Two-mode scanning probe microscopy}
In order to use the NW as a scanning probe, we approach the sample and
scan it in a plane below the NW tip.  By monitoring the NW's
mechanical properties, i.e.\ the frequency, dissipation, and
orientation of its doublet modes, we image the sample topography via
the tip-sample interaction.  Such microscopy can be accomplished by
measuring the NW thermal displacement spectral density as the sample
surface is scanned below it.  Although such a measurement provides a
full mechanical characterization of the modes, it is time-consuming
due to the small displacement.  A technique more amenable to fast
spatial scans uses the resonant excitation of the doublet modes
through two independent phase-locked loops (PLLs) to track both
frequencies simultaneously (see Appendix).

Fig.~2 shows the frequency shifts $\Delta f_1 (x, y)$ and $\Delta f_2
(x, y)$ of the doublet modes as a sample is scanned below the tip of
NW1.  The sample consists of nine 5-$\mu$m long and 200-nm thick
finger gates of Au on a Si substrate, radially disposed and equally
spaced along a semicircle (see Supplementary Information for an SEM of
the sample).  The finger gates and their tapered shape are intended to
provide edges at a variety of different angles, highlighting the
directional sensitivity of the orthogonal modes.  The measurement in
Fig.~2 is performed using the PLL with an oscillation amplitude of 50
nm at a distance of 70 nm from the Au surfaces in ``open loop'', i.e.\
without feedback to maintain a constant distance from the sample.  The
Au gates are grounded during the measurement.

The frequency shift images clearly delinate the topography of the
patterned sample, with each mode showing stronger contrast for
features aligned along orthogonal directions.  These two directions
(identified by noting the direction of the fingers with maximum
contrast) agree with the angle $\theta_0 = 15^\circ$ measured for NW1
via the thermal noise shown in Fig.~1(d).  Edges, i.e.\ large
topographical gradients, pointing perpendicular (parallel) to the mode
oscillation direction appear to produce the strongest (weakest)
contrast.  Tip-sample interactions producing the frequency shifts in
non-contact AFM can include electrostatic, van der Waals, or chemical
bonding forces depending on the distance.  In our case, because of the
large spacing, they are dominated by electrostatic forces.

\section{Extraction of spatial tip-sample force derivatives}
In order to understand the measurements, we describe the motion of the
NW tip in each of the two fundamental flexural modes as a driven
damped harmonic oscillator:
\begin{equation}
  m \ddot{r_i} + \Gamma_i \dot{r_i} + k_i r_i = F_{th} + F_i,
  \label{eq1}
\end{equation}
where $m$ is the effective mass of the fundamental flexural modes,
$F_{th}$ is the Langevin force, $F_i$ is the component of the
tip-sample force along $\mathbf{\hat{r}}_i$, and $i = 1, 2$.
Following the treatment of Gloppe et
al.~\cite{gloppe_bidimensional_2014} and expanding $F_i$ for small
oscillations around the equilibrium $r_i = 0$, we have $F_i \approx
F_i(0) + \left. r_j \partial F_i / \partial r_j\right |_{0}$.  By
replacing this expansion in (\ref{eq1}), we find that this tip-sample
interaction produces new doublet eigenmodes with modified spring
constants $k_i'$ (see Supplementary Information):
\begin{equation} \label{eq2}
\begin{split}
  k_{1,2}' &= \frac{1}{2} \bigg [k_1 + k_2 - F_{11} - F_{22}\\
  &\qquad \pm \sqrt{(k_1 - k_2 - F_{11} + F_{22})^2 + 4 F_{12} F_{21}}
  \bigg ], \\
\end{split}
\end{equation}
where we use a shorthand notation for the force derivatives $F_{ij}
\equiv \left. \partial F_i / \partial r_j \right |_{0}$.  In addition
to inducing frequency shifts, tip-sample force derivatives with
non-zero shear components ($F_{ij} \neq 0$ for $i \neq j$) couple the
two flexural modes and rotate their oscillation direction along two
new basis vectors
$\mathbf{\hat{r}}_i'$~\cite{faust_nonadiabatic_2012}.  These new modes
remain orthogonal for conservative force fields (i.e.\
$F_{12}=F_{21}$), but lose their orthogonality for non-conservative
force fields.

For tip-sample force derivatives that are much smaller than the bare
NW spring constant -- which is the case here -- the modified spring
constants $k_i' \approx k_i - F_{ii}$.  Therefore, by monitoring the
frequency shift between the bare resonances and those modified by the
tip-sample interaction $\Delta f_i = f_i' - f_i$, we measure:
\begin{equation}
  \left. \frac{\partial F_i}{\partial r_i} \right |_0 \approx -2 k_i \left ( \frac{\Delta f_i}{f_i} \right ),
\label{eq3}
\end{equation}
Because the oscillation amplitude ($\sim 50$ nm) is small compared to
the tip diameter ($\sim 350$ nm), we can ignore variations of the
force derivative over the oscillation cycle.  Just as in conventional
1D dynamic lateral force microscopy, each $\Delta f_i(x, y)$ depends
on the derivative $F_{ii}$, i.e.\ on the force both projected and
differentiated along the mode oscillation direction.  The NW mode
doublet, however, is able to simultaneously measure force derivatives
along orthogonal directions.

In the same limit of small derivatives considered for (\ref{eq3}), the
rotation of the mode axes reveals the shear, or cross derivatives of
the force, through,
\begin{equation}
  \left. \frac{\partial F_i}{\partial r_j} \right |_0 \approx \left | k_i - k_j \right | \tan \phi_i,
\label{eq4}
\end{equation}
for $i \neq j$ and where $\phi_i = \theta_i-\theta_0$ is the angle
between the mode direction $\mathbf{\hat{r}}_i'$ in the presence of
tip-surface interaction and the bare mode direction
$\mathbf{\hat{r}}_i$.  Therefore by monitoring the doublet mode
frequency shifts and oscillation directions, we can completely
determine the in-plane tip-sample force derivatives $F_{ij}$.

\begin{figure}[b]
  \centering
  \includegraphics[width=1\columnwidth]{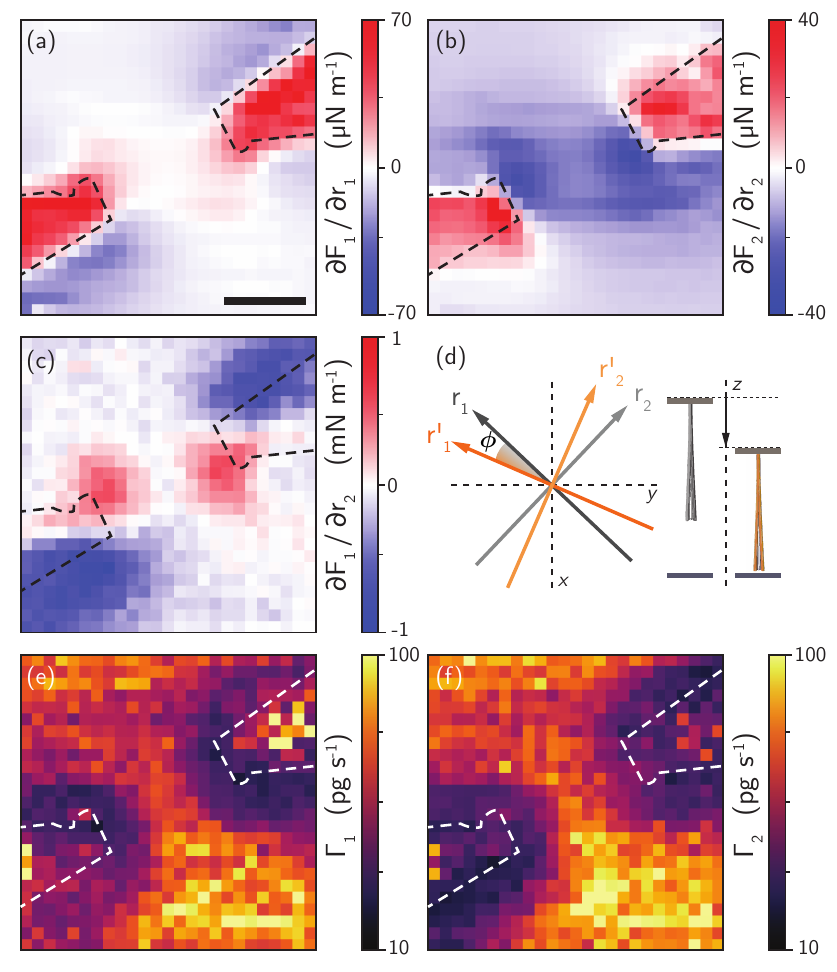}
\caption{
\label{fig3}
\textbf{2D Force derivative and dissipation images}. \textbf{a, b, c},
Force derivatives $F_{11}$, $F_{22}$, and $F_{12} = F_{21}$,
respectively, extracted from thermal noise data as a function of the
$xy$-position of NW2 70 nm over the gated sample.  Overlayed dashed
lines indicate the edges of the finger gates, as obtained by SEM.
\textbf{d}, Schematic picture of the direction of the flexural
eigenmodes at two tip-sample distances $z$.  The black and gray axes
indicate the direction of the unperturbed axes with $\theta_0 =
46^\circ$.  The eigenmodes rotate through an angle $\phi$ when brought
closer to the sample.  \textbf{e, f} Dissipation $\Gamma_i$ of each
mode as a function of the $xy$-position.}
\end{figure}

\section{Imaging static in-plane force derivatives and dissipation}
A complete measurement of static in-plane force derivatives and
dissipations is shown in Fig.~3 for a section of the ``finger'' sample
using a second NW, NW2.  $\Delta f_i$ and $\Gamma_i$ are extracted
from fits to the spectral density of the thermal noise measured as the
sample was scanned below the NW at a fixed spacing of 70 nm.  By
assuming a conservative tip-sample interaction ($F_{ij}=F_{ji}$), we
determine $\phi = \phi_1 = \phi_2$ from the frequency and integrated
power of each mode in the spectral density compared to a similar
measurement far from the surface, giving $\theta_0 = 46^\circ$.  Using
these data along with (\ref{eq3}) and (\ref{eq4}), we produce maps of
$F_{ij}(x, y)$ and $\Gamma_i(x, y)$.  The measurements show strong
positive followed by negative $F_{ii}$ for edges perpendicular to the
NW mode oscillation $\mathbf{\hat{r}}_i$.  In non-contact AFM,
tip-samples forces are generally attractive and become more so with
decreasing tip-sample distance and increased interaction area.  As the
NW approaches a Au edge perpendicular to $\mathbf{\hat{r}}_i$ from a
position above the lower Si surface, it will experience an
increasingly attractive force, i.e.\ a positive $F_{ii}$.  After the
midpoint of the tip crosses this edge, the attractive force starts to
drop off, resulting in a negative $F_{ii}$.  As expected, measurements
show that $F_{ij}$ for $i \neq j$ are peaked around curved edges.
Measured tip-sample dissipations $\Gamma_i$ are nearly isotropic and
appear to reflect the different materials and tip-sample spacings over
the Au fingers and the Si substrate.  Similar 1D measurements of
non-contact friction also show lower values over conducting surfaces
like Au compared to insulating surfaces like Si and point to charge
fluctuations as the origin for the
dissipation~\cite{stipe_noncontact_2001,kuehn_dielectric_2006}.

\begin{figure*}[t]
  \centering
  \includegraphics[width=2\columnwidth]{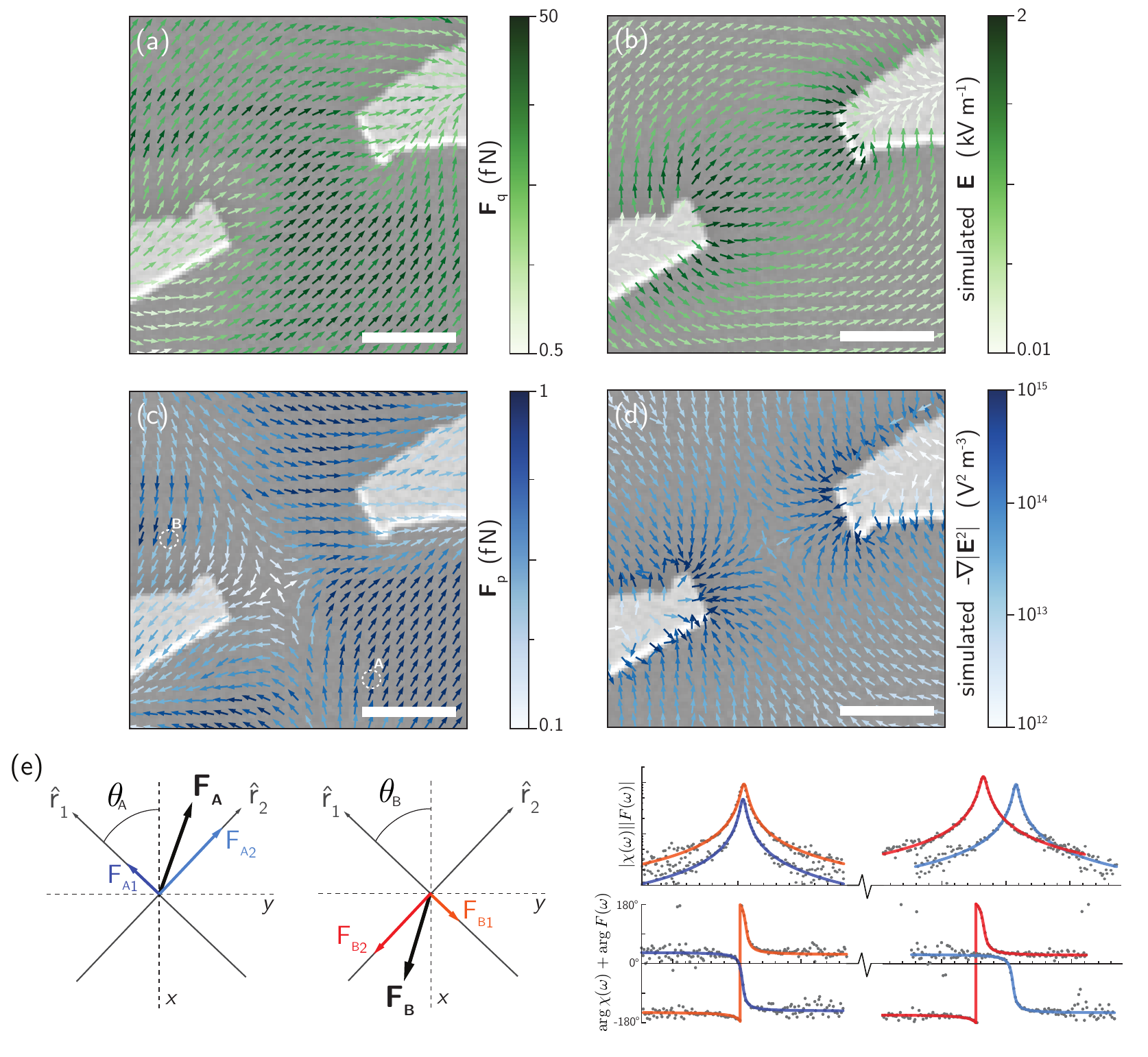}
\caption{
\label{fig4}
\textbf{Vector plots of electrostatic force fields}.  \textbf{a}
Vector plot of the force field induced by an electric field on the
charged NW.  The vector plot is overlayed on top of a SEM image of the
sample.  \textbf{b}, Vector plot of the simulated electric field
$\mathbf{E}(\mathbf{r})$ induced by the biased finger gates.
\textbf{c} Vector plot of the force field induced by the electric
field on the polarizable NW.  The voltage amplitude applied to the
gates in \textbf{a} and \textbf{c} is $V = 2$ and $20$ mV,
respectively, the vertical spacing between NW2 and the Au gates is 70
nm, and $\theta_0 = 46^\circ$.  \textbf{d}, Vector plot of the
simulated values of $-\mathbf{\nabla}|\mathbf{E(\mathbf{r})}|^2$
induced by the biased finger gates.  \textbf{e}, The diagram shows
schematic diagrams of how force vectors are extracted from
measurements at the example positions A and B.  The amplitude and
orientation of the force is extracted from fits to the amplitude and
phase response as a function of frequency of the two modes, shown in
the corresponding plots.}
\end{figure*}

\section{Imaging a dynamic in-plane force field}
In-plane force fields can also be mapped with a thermally limited
sensitivity of 5 aN/$\sqrt{\text{Hz}}$ by measuring the driven NW
response, as shown in Fig.~4.  The frequency of the force to be
measured must be swept through the NW doublet resonances, while the
amplitude and phase of the displacement response is recorded.  As an
example, we apply a small AC voltage $V \cos(2 \pi f t)$ with
frequency $f$ producing an oscillating electric field
$\mathbf{E}(\mathbf{r}, t)$ between opposing finger gates.  In
scanning the NW and measuring its response, we distinguish between two
types of forces.  The first shows a linear dependence on the electric
field strength and is associated with charge: $\mathbf{F_q} = q
\mathbf{E}$, where $q$ is the net charge on the NW tip.  The second
force exhibits a quadratic dependence on the field strength and is
associated with the induced dipolar moment of the dialectric NW, i.e.\
with its polarizability: $\mathbf{F_p} = -\mathbf{\nabla} (\alpha
|\mathbf{E}|^2)$, where $\alpha$ is the effective polarizability of
the GaAs/AlGaAs NW~\cite{rieger_frequency_2012}.  Due to their linear
and quadratic dependence on $\mathbf{E}$, respectively, $\mathbf{F_q}$
drives the NW at frequency $f$, while $\mathbf{F_p}$ drives it at DC
and $2f$.  As a result, the two interactions can be spectrally
separated.

The magnitude and orientation of the driving force along each doublet
mode direction can be extracted from the displacement and phase
response of each mode as a function of frequency, as shown
schematically in Fig.~4(e).  A measurement of the thermal noise
spectrum in the absence of the AC drive is used to calibrate the
orientation of the mode doublet.  By scanning the sample in a plane 70
nm below NW2 and measuring thermal motion and driven response at each
point, we construct vectorial maps of the $\mathbf{F}_q$ and
$\mathbf{F}_p$, as shown in Fig.~4(a) and (c) and described in the
Appendix.  These measured force fields are compared to the
$\mathbf{E}$ and $-\mathbf{\nabla}|\mathbf{E}^2|$ fields simulated by
FEM analysis (COMSOL) using the real gate geometries in Figs.~4(b) and
(d).  Dividing the measured force fields by the corresponding
simulations, we determine the net charge on the tip of NW2 to be $q =
30 \pm 10$ $e$, where $e$ is the fundamental charge, and the effective
polarizability $\alpha = 10^{-29}$ $\text{C}/(\text{V} \cdot
\text{m})$, which is roughly consistent with the size and dielectric
constant of the NW.  The ability to vectorially map electric fields on
the nanometer-scale extends the capability of conventional AFM to
image
charges~\cite{stern_deposition_1988,schonenberger_observation_1990}
and contact potential differences~\cite{nonnenmacher_kelvin_1991} and
has applications in localizing electronic defects on or near surfaces,
e.g.\ in microelectronic failure analysis.

\section{Conclusion and Outlook}
 Our measurements demonstrate the potential of NWs as
sensitive scanning vectorial force sensors.  By monitoring two
orthogonal flexural modes while scanning over a sample surface, we map
forces and force derivatives in 2D.  While we demonstrate the
technique on electrostatic tip-sample interactions, it is generally
applicable and could, for instance, be used to measure magnetic forces
with proper functionalization of the NW tip or even in
liquid~\cite{sanii_high_2010} for the study of batteries, water
splitting, or fuel cells.  Note that in the presented data, the 350-nm
diameter of the NW tip strongly limits the spatial resolution of the
microscopy.  Nevertheless, the same technique could be applied to NWs
grown or processed to have sharp tips, presenting the possibility of
atomic-scale or even sub-atomic-scale force microscopy with
directional sensitivity, a feat already achieved in
1D~\cite{kawai_ultrasensitive_2010}.  Therefore, one could imagine the
use of vectorial NW-based AFM of tip-sample forces and non-contact
friction to reveal, for example, the anisotropy of atomic bonding
forces.

\section{Acknowledgements}

\acknowledgements We thank Sascha Martin and the mechanical workshop
at the University of Basel Physics Department for help in designing
and building the NW microscope and Jean Teissier for useful
discussions.  We acknowledge the support of the ERC through Starting
Grant NWScan (Grant No. 334767), the Swiss Nanoscience Institute
(Project P1207), and the Swiss National Science Foundation (Ambizione
Grant No. PZOOP2161284/1 and Project Grant No. 200020-159893).

\appendix

\section{Appendix A: Nanowire growth and processing}
 MBE synthesis of the GaAs/AlGaAs NWs starts on a Si
substrate with the growth of a 290-nm thick GaAs NW core along
$[1\,\bar{1}\,1]$ by the Ga-assisted method detailed in Uccelli et
al.~\cite{uccelli_three-dimensional_2011} and Russo-Averchi et
al~\cite{russo-averchi_suppression_2012}.  Axial growth is stopped
once the NWs are about 25-$\mu$m long by temporarily blocking the Ga
flux and reducing the substrate temperature from 630 down to
$465^{\circ}$C.  Finally, a 50-nm thick Al$_{0.51}$Ga$_{0.49}$As shell
capped by a 5-nm GaAs layer is grown~\cite{heigoldt_long_2009}.  For
optimal inferometric detection of the NWs, we focus on NWs within 40
$\mu$m from the substrate edge.  To avoid measuring multiple NWs or
having interference between NWs, we reduce the NW density in the area
of interest using a micromanipulator under an optical microscope.

\section{Appendix B: Nanowire positioning and displacement detection}
 The measurement chamber includes two stacks of piezoelectric
positioners (Attocube AG) to independently control the 3D position of
the NW cantilevers and the sample of interest with respect to the
fixed detection optics.  The top stack is used to align a single NW
within the focus of a fiber-coupled optical interferometer used to
detect its mechanical motion~\cite{rugar_improved_1989}.  Once the NW
and interferometer are aligned, the bottom stack is used to approach
and scan the sample of interest with respect to the NW cantilever.
Light from a laser diode with wavelength of 635 nm is sent through one
arm of a 50:50 fiber-optic coupler and focused by a pair of lenses to
a 1-$\mu$m spot.  The incident power of $\sim5 \mu$W does not
significantly heat the NW as confirmed by measurements of laser power
dependence and mechanical thermal motion.  Despite the sub-$\mu$m
diameter of the NW, it reflects a portion of the light back into the
fiber, which interferes with light reflected by the fiber's cleaved
end.  The resulting low-finesse Fabry-Perot interferometer acts as a
sensitive sensor of the NW displacement, where the interference
intensity is measured by a fast photoreceiver with an effective 3 dB
bandwidth of 800 kHz.

\section{Appendix C: Measurement protocols}
Measurements of NW thermal motion used for Figs.~1, 3 and 4 are made
using a analog-to-digital converter (National Instruments).  Fits
based on the thermally driven damped harmonic oscillator of
(\ref{eq1}) are used to extract $\theta_0$, $f_i$, $\Gamma_i$, $m$,
and $\phi_i$.  Faster measurements, as in Fig.~2, are performed by
resonantly exciting both flexural modes using the piezoelectric
transducer.  Due to the high $Q_i$ and large $\delta$ of the
resonances, it is possible to monitor and control both modes
simultaneously.  As in standard AFM, we use a lock-in amplifier
(Zurich Instruments UHFLI) to resonantly drive each mode and
demodulate the resulting optical signal measured by the photoreceiver.
For each mode, its shift in resonant frequency $\Delta f_i$ is tracked
by a PLL, while proportional-integral control of the excitation
voltage maintains a constant oscillation amplitude.  Measurements of
the NW amplitude and phase response to a time-varying gate voltage
applied to the the sample, as shown in Fig.~4, are made using the same
lock-in.  Fits to the amplitude and phase response of each mode as a
function of frequency are based on the mechanical susceptibility
$\chi_i(\omega) = 1/(k_i - m \omega^2 + i \Gamma_i \omega )$ of a
damped harmonic oscillator, where $\omega = 2 \pi f$.  The oscillation
amplitude $|\chi_i(\omega)||F_i(\omega)|$ excited for mode $i$ gives
the magnitude of the driving force $F_i$ along $\mathbf{\hat{r}}_i$.
The phase $\arg F_i(\omega)$ gives the orientation of $F_i$ along
either positive or negative $\mathbf{\hat{r}}_i$.

\clearpage

\begin{figure*}[htb]
  \centering
  \includegraphics[width=2\columnwidth]{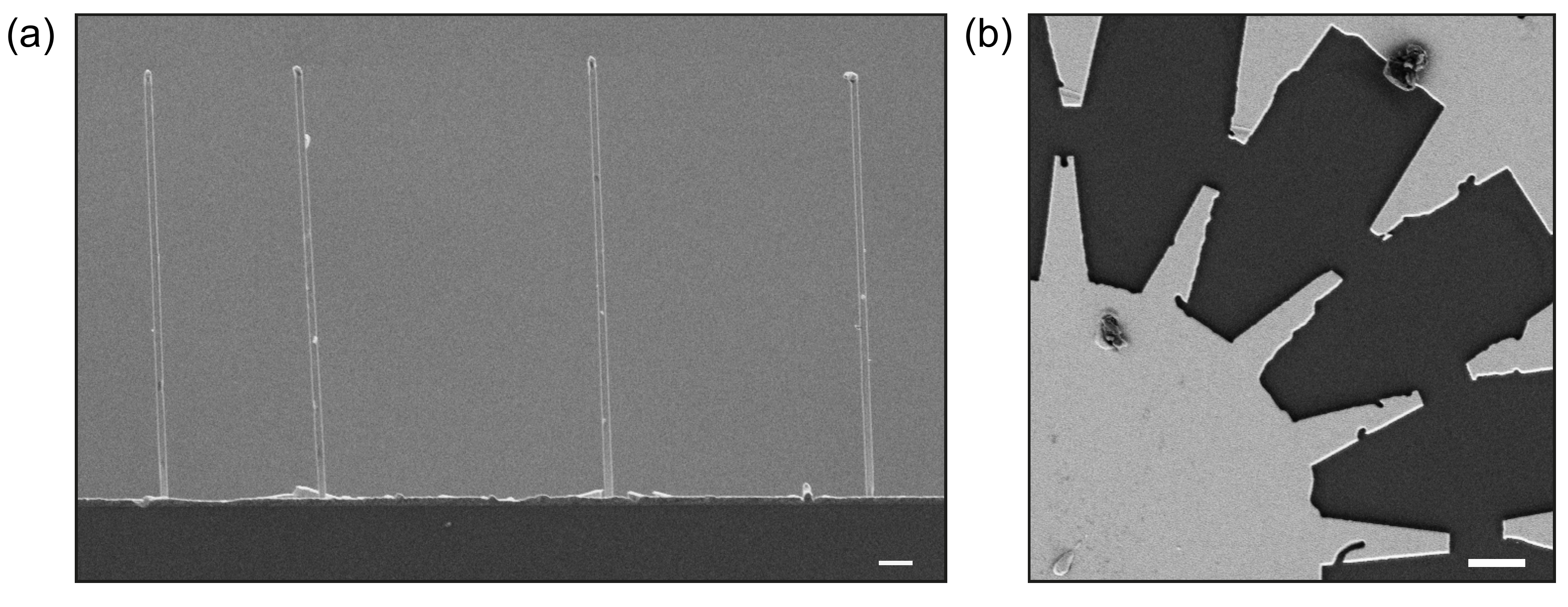}
  \label{figSup}
  \caption{\textbf{Supplementary}. \textbf{a}, Zoomed out SEM image of
    the edge of the NW sample, showing several similar NWs. \textbf{b}
    Zoomed out SEM image of the finger gate sample.  The scalebar in
    both figures corresponds to 2$\mu$m.}
\end{figure*}

\section{Supplementary Information: Nanowire motion in presence of
  tip-sample force}

We describe the motion of the NW tip in each of the two fundamental
flexural modes as a driven damped harmonic oscillator:
\begin{equation}
  m \ddot{r_i} + \Gamma_i \dot{r_i} + k_i r_i = F_{th} + F_i,
  \label{eq1}
\end{equation}
where $m$ is the effective mass of the fundamental flexural modes,
$F_{th}$ is the Langevin force, $F_i$ is the component of the
tip-sample force along the mode oscillation direction
$\mathbf{\hat{r}}_i$, and $i = 1, 2$.  Following the treatment of
Gloppe et al.\ \cite{gloppe_bidimensional_2014} and expanding $F_i$
for small oscillations around the equilibrium $r_i = 0$, we have,
\begin{equation}
  F_i \approx F_i(0) + \left. r_j \frac{\partial F_i}{\partial r_j} \right |_{0}.
\label{eq1.5}
\end{equation}
By replacing this expansion in (\ref{eq1}), we have,
\begin{equation}
  m \ddot{r_i} + \Gamma_i \dot{r_i} + k_i r_i = F_{th} + F_i(0) + F_{ij} r_j,
  \label{eq2}
\end{equation}
where we use a shorthand notation for the force derivatives $F_{ij}
\equiv \left. \frac{\partial F_i}{\partial r_j} \right |_{0}$.  As
(\ref{eq2}) makes clear, the derivatives of the tip-sample force modify
and couple the NW's two unperturbed flexural modes.  We now rewrite
this equation in vectorial form as,
\begin{equation}
  m \ddot{\mathbf{r}} + \bar{\Gamma} \cdot \dot{\mathbf{r}} + \bar{K} \cdot \mathbf{r} = F_{th} + \mathbf{F}_0,
  \label{eq3}
\end{equation}
where $\mathbf{r} = \left ( \begin{smallmatrix} r_1 \\
    r_2 \end{smallmatrix} \right )$ and the equilibrium tip-sample
force $\mathbf{F}_0 = \mathbf{F}(\mathbf{r} = 0)$.  The dissipation
and spring constant matrices are defined by:
\begin{gather}
  \bar{\Gamma} \equiv \begin{pmatrix} \Gamma_1 & 0 \\ 0 & \Gamma_2 \end{pmatrix},\label{eq4}\\[2ex]
  \bar{K} \equiv \begin{pmatrix} k_1 - F_{11} & -F_{21} \\ -F_{12} &
    k_2 - F_{22} \end{pmatrix},
  \label{eq5}
\end{gather}
where the role of the shear cross-derivatives, i.e.\ $F_{ij}$ for $i
\neq j$, in coupling the unperturbed NW modes is
clear~\cite{faust_nonadiabatic_2012}.  In the presence of weak
tip-surface interactions, as those studied here, the dissipation rates
of the NW modes are neglibly small compared to their unperturbed
resonant frequencies, i.e.\ $\frac{\Gamma}{2 m} \ll
\sqrt{\frac{k}{m}}$.  In this limit, the NW mode frequencies and
oscillation directions are determined by $\bar{K}$.  Therefore, by
diagonalizing $\bar{K}$, we find a new pair of uncoupled flexural
modes.  The corresponding spring constants and mode directions have
been modified from the unperturbed state by the spatial tip-sample
force derivatives $F_{ij}$:
\begin{align}
  k_1' &= \frac{1}{2} \bigg [k_1 + k_2 - F_{11} - F_{22} \nonumber \\
  &\qquad + \sqrt{(k_1 - k_2 - F_{11} + F_{22})^2 + 4 F_{12} F_{21}}
  \bigg ],\label{eq6}\\
  \mathbf{\hat{r}}_1' &= \frac{1}{\sqrt{(k_2 - F_{22} - k_1')^2 +
      F_{12}^2}} \begin{pmatrix} k_2 - F_{22} - k_1' \\
    F_{12} \end{pmatrix};\label{eq7}\\[4ex]
  k_2' &= \frac{1}{2} \bigg [k_1 + k_2 - F_{11} - F_{22} \nonumber \\
  &\qquad - \sqrt{(k_1 - k_2 - F_{11} + F_{22})^2 + 4 F_{12} F_{21}} \bigg ],\label{eq8}\\
  \mathbf{\hat{r}}_2' &= \frac{1}{\sqrt{(k_1 - F_{11} - k_2')^2 +
      F_{21}^2}} \begin{pmatrix} F_{21} \\ k_1 - F_{11} -
    k_2' \end{pmatrix}.
  \label{eq9}
\end{align}
These new modes remain orthogonal ($\mathbf{\hat{r}}_1' \cdot
\mathbf{\hat{r}}_2' = 0$) for conservative force fields ($\nabla
\times \mathbf{F} = 0$, i.e.\ $F_{12} - F_{21} = 0$), but lose their
orthogonality for non-conservative force fields.

For tip-sample force derivatives that are much smaller than the bare NW
spring constants -- which is the case here -- the modified spring
constants and the modified mode directions in (\ref{eq6}) -
(\ref{eq9}) can be approximated to first order in the derivatives:
\begin{gather}
  k_1' \approx k_1 - F_{11}, \label{eq10}\\
  \mathbf{\hat{r}}_1' \approx \frac{1}{\sqrt{(k_1 - k_2)^2 +
      F_{12}^2}} \begin{pmatrix} k_1 - k_2 \\
    -F_{12} \end{pmatrix};\label{eq11}\\[4ex]
  k_2' \approx k_2 - F_{22}, \label{eq12}\\
  \mathbf{\hat{r}}_2' \approx \frac{1}{\sqrt{(k_1 - k_2)^2 +
      F_{21}^2}} \begin{pmatrix} F_{21} \\ k_1 - k_2 \end{pmatrix}.
  \label{eq13}
\end{gather}
In the limit of small dissipation discussed previously, the
unperturbed resonance frequencies of the flexural modes are given
by $f_i = \frac{1}{2 \pi} \sqrt{\frac{k_i}{m}}$.  Similarly, the
modified resonance frequencies are given by $f_i' = \frac{1}{2 \pi}
\sqrt{\frac{k_i'}{m}}$.  For small tip-sample force derivatives, we
apply (\ref{eq10}) and (\ref{eq12}) and arrive at $f_i' \approx
\frac{1}{2 \pi} \sqrt{\frac{k_i - F_{ii}}{m}}$.  Expanding to first
order in $F_{ii}$, we find $f_i' \approx f_i - \frac{f_i}{2 k_i}
F_{ii}$.  Solving in terms of the frequency shift induced by the
tip-sample interaction, we have:
\begin{equation}
  \Delta f_i = f_i' - f_i \approx - \frac{f_i}{2 k_i} F_{ii}.
\label{eq14}
\end{equation}
We can now write a relation for $F_{ii} = \partial F_i / \partial r_i$
in terms of the measured frequency shift induced by the tip-sample
interaction and properties of the unperturbed NW modes:
\begin{equation}
  \frac{\partial F_i}{\partial r_i} \approx - 2 k_i \left (\frac{\Delta f_i}{f_i} \right ).
\label{eq15}
\end{equation}
Using (\ref{eq11}) and (\ref{eq13}), we can also write an expression
involving the angle $\phi_i$ between the bare mode direction
$\mathbf{\hat{r}}_i$ and the corresponding modified mode direction
$\mathbf{\hat{r}}_i'$:
\begin{equation}
  \tan{\phi_i} \approx \frac{F_{ij}}{\left | k_i - k_j \right |}.
\label{eq16}
\end{equation}
This equation then allows us to solve for $F_{ij} = \partial F_i
/ \partial r_j$ for $ i \neq j$ in terms of the measured angles
$\phi_i$ and the unperturbed spring constants:
\begin{equation}
  \frac{\partial F_i}{\partial F_j} \approx | k_i - k_j | \tan{\phi_i}.
\label{eq17}
\end{equation}
In this way, we are able to measure all spatial tip-sample force derivatives in
the small interaction limit (i.e.\ all derivatives much smaller than the
unperturbed spring constants).

\end{document}